\documentstyle[11pt,Cargesepasp,twoside,epsf,psfig]{article}
\def\lsim   {{_{<}\atop^{\sim}}}
\def\gsim   {{_{>}\atop^{\sim}}}

\def\edcomment#1{\iffalse\marginpar{\raggedright\sl#1\/}\else\relax\fi}
\reversemarginpar

\begin{document}
\title{The Onset of Cluster Formation around Intermediate Mass Stars}
 \author{Leonardo Testi, Francesco Palla \& Antonella Natta}
\affil{Osservatorio Astrofisico di Arcetri, Largo E. Fermi 5, 
Firenze, Italy}

\begin{abstract}
We present the results of a recent near-infrared survey of the fields
surrounding a large sample of intermediate-mass pre-main-sequence
(Herbig~Ae/Be) stars. While late-type Be and Ae stars are never associated
with conspicuous groups of young lower mass stars, early-type Be stars are
usually found within rich clusters. This finding has been tested against
possible biases due to different mass sensitivity of the observations or
dynamical dissipation of the clusters around older stars.  Our results
suggest that massive stars are preferentially produced in dense stellar
clusters, possibly by dynamical interaction rather than by standard gas
accretion as in the case of lower mass stars. The possibility that the
observed correlation between maximum stellar mass and cluster richness could
be the results of random sampling the cluster size spectrum and the stellar
IMF is also discussed.  Future observational tests capable of discriminating
between these two competitive models are outlined.  \end{abstract}

\section{Introduction}

It is now well established by means of 
direct and indirect observations
that most, if not all, stars are formed
in groups rather than in isolation
(Clarke, Bonnell \& Hillenbrand~2000; Carpenter~2000).
An important result that strongly
constrains theories of massive stars and stellar clusters formation
is that the stellar density of young stellar clusters seems to depend
on the most massive star in the cluster.
Low-mass stars are usually found to form in loose groups with
typical densities of a few stars per cubic parsec (Gomez et al.~1993),
while high-mass stars are found in dense clusters of up to 10$^4$ stars per
cubic parsec (see e.g. Hillenbrand \& Hartmann~1998).
The transition between these two modes of formation should occur in the 
intermediate-mass regime, namely 2$\lsim{\rm M/M_\odot}\lsim$15.

Here, we present the results of an extensive near infrared (NIR) survey for
young clusters around intermediate-mass stars aimed at the detection and
characterization of their clustering properties (Testi et al. 1997; 1998;
1999). We were primarily motivated by the expectation that at NIR
wavelengths, particularly in the K-band (2.2~$\mu$m), the reduced extinction
would allow the detection of embedded young stars in the vicinity of the
intermediate-mass stars.


The selected sample consists of 44 intermediate-mass pre-main sequence stars 
(Herbig Ae/Be stars) taken from the catalogues of Finkenzeller \& Mundt~(1984)
and Th\'e et al.~(1994). The
primary selection criterion was to cover as uniformly as possible the 
O9--A7 spectral type range. As an additional constraint, we required the sources
to be observable from the northern hemisphere. There appear to be no particular
biases in our sample selection, except for the fact that Ae stars tend to be 
closer to the Sun than Be stars (Testi et al.~1998).
Herbig Ae/Be stars themselves are optically visible stars
with relatively low extinction (typically A$_{\rm V}<5$-10~mags), and the
molecular material of the parental cloud cores has been already dissipated
in many cases (Fuente et al.~1998). Thus, we expect that the star formation
event that produced
the Herbig star, and the possible accompaining population of lower mass stars,
should be terminated in most of the regions in our sample.

\begin{figure}
\centerline{\psfig{figure=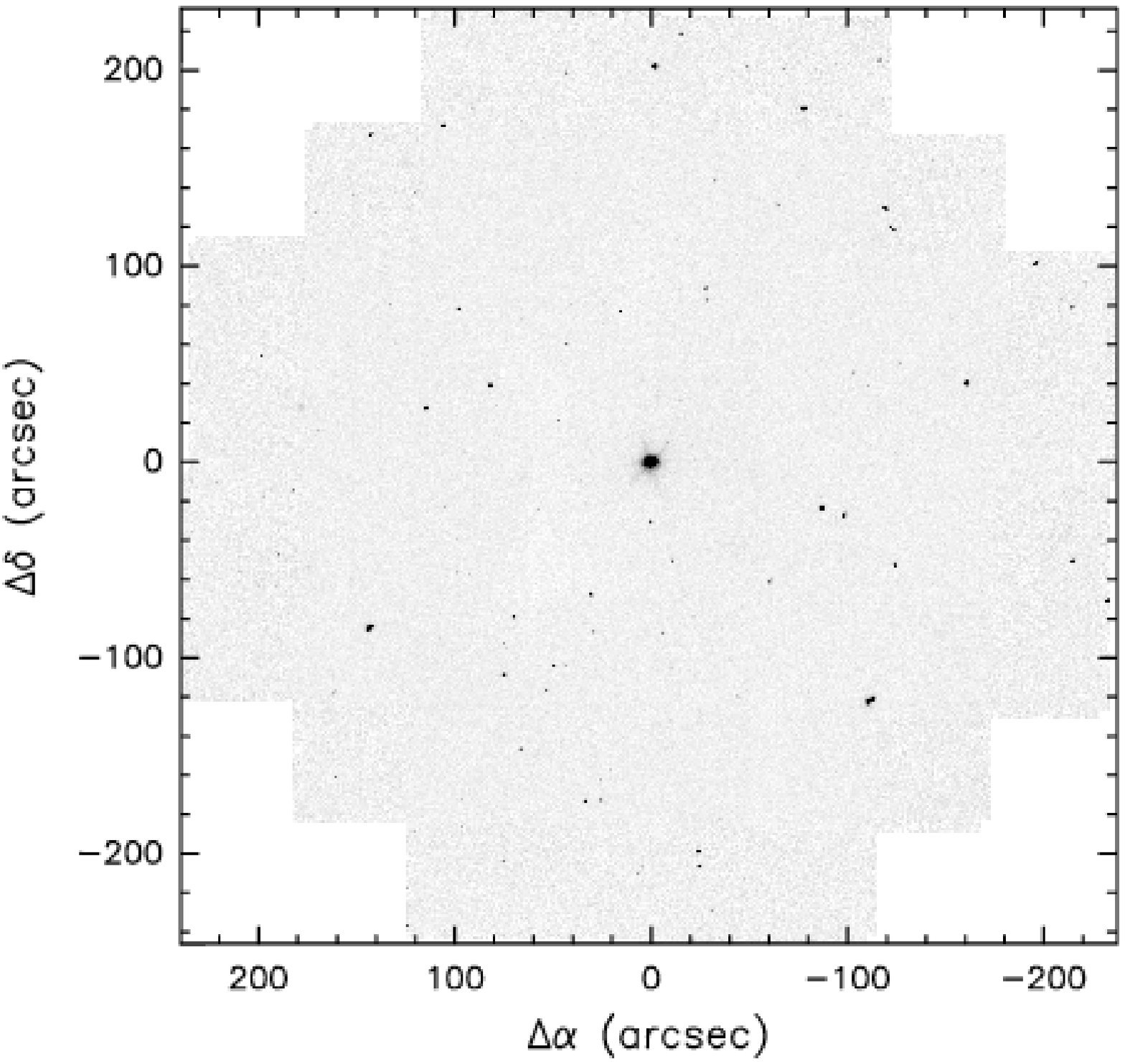,height=6.5cm,angle=-90}
            \hskip 0.5cm
            \psfig{figure=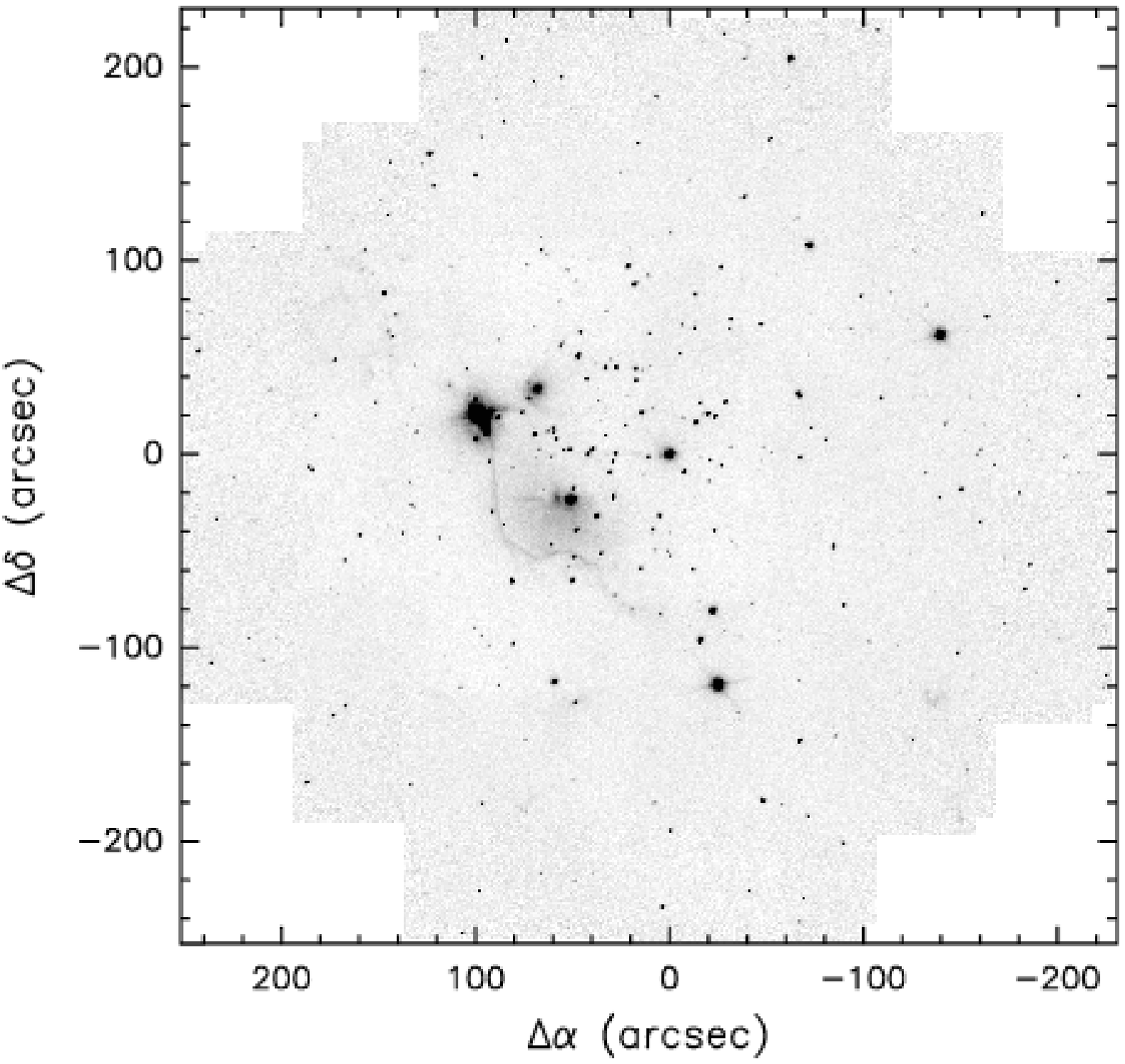,height=6.5cm,angle=-90}}
\centerline{\psfig{figure=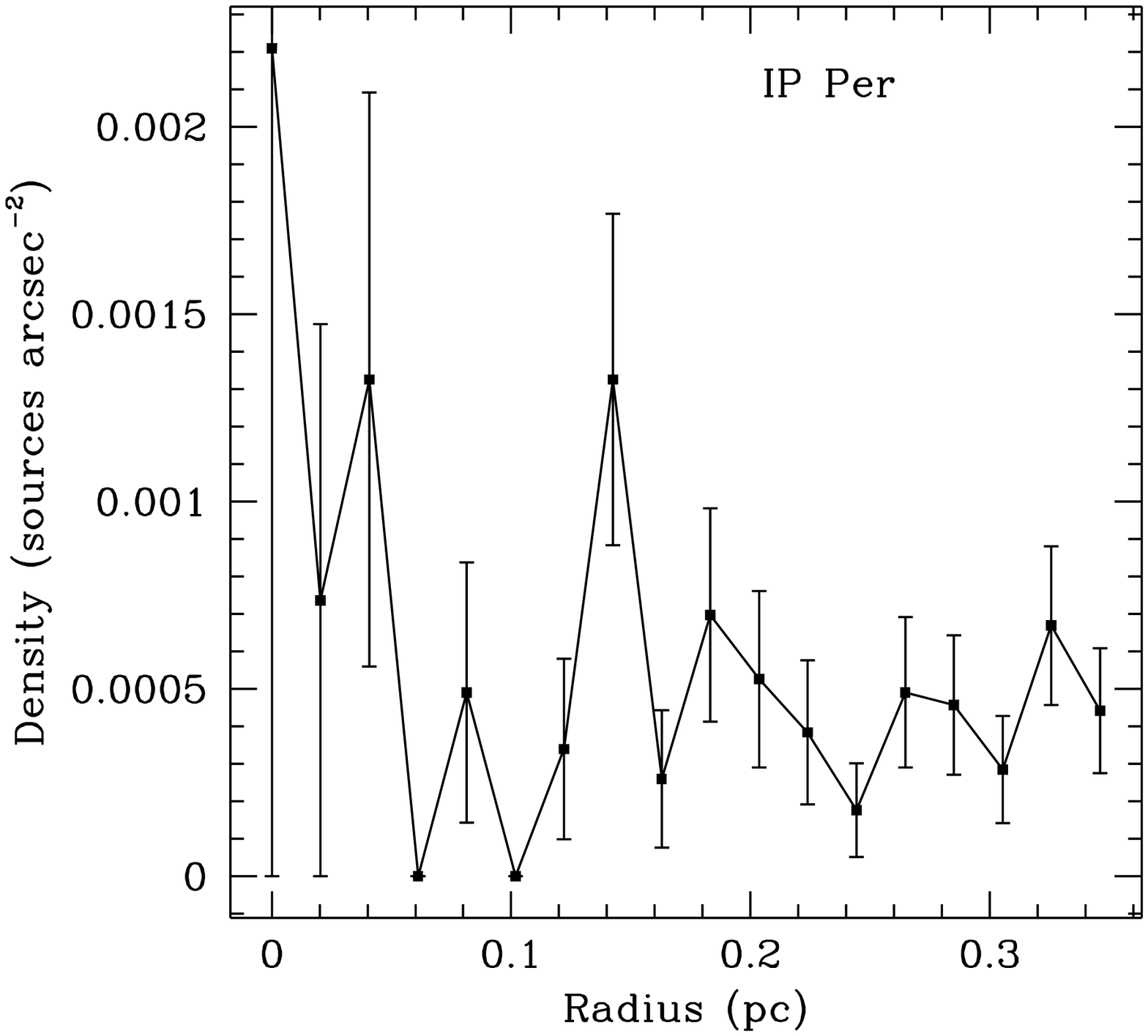,height=5.8cm,angle=0}
            \hskip 0.5cm
            \psfig{figure=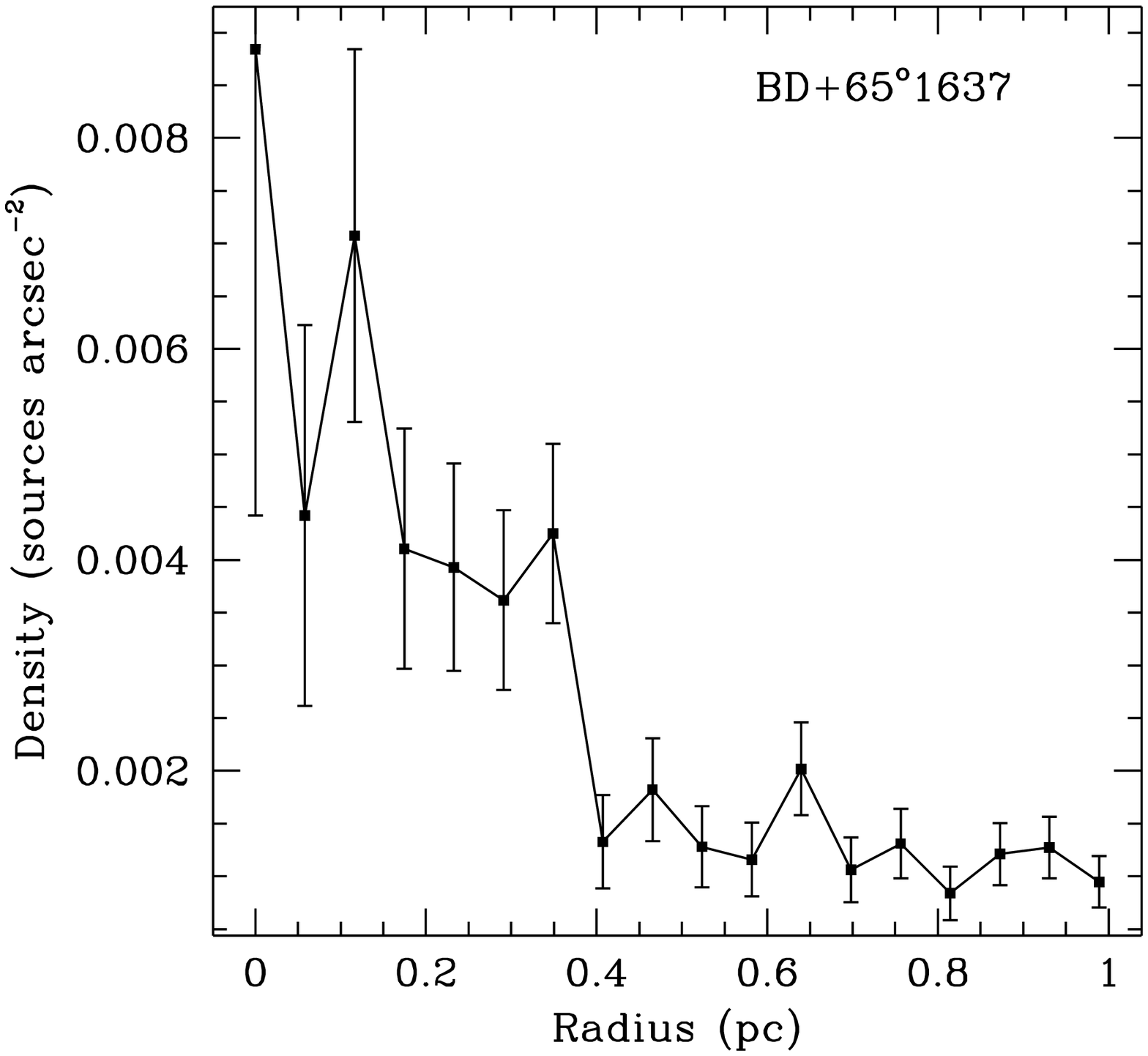,height=5.8cm,angle=0}}
\caption{\label{fk}K-band images (top panels) and K-band radial
stellar surface density
profiles (bottom panels) of the fields surrounding IP~Per (A3)
and BD$+$65$^\circ$1637 (B2)}
\end{figure}
The fields surrounding the target stars were observed with the Arcetri NIR 
camera (ARNICA) mounted either at the TIRGO or at the NOT telescopes.
The relatively large field of view and sensitivity of ARNICA allowed us
to obtain wide and deep K-band images of the regions around the target stars.
In Figure~\ref{fk} we show two fields surrounding 
IP~Per (spectral type A3e) and BD$+$65$^\circ$1637 (B2e).
Given the distance
and expected ages of these two targets, our completeness absolute magnitudes at
K-band correnspond to a minimum detectable stellar mass of less than
0.1~M$_\odot$.
The method used to convert observational magnitudes into stellar masses
via pre-main sequence evolutionary tracks is
described in Testi et al.~(1998), together with the adopted calibrations.

\section{Survey Results}

Testi et al.~(1997) discussed several different quantitative
{\it richness indicators} to study the clustering properties of the 
young stellar populations around Herbig Ae/Be stars. The most
effective method to detect clusters removing the field stars
contamination was found to be the study of the K-band radial stellar surface
density profiles centered on the Herbig stars.
As an example, in Figure~\ref{fk} we show the profiles
for the fields surrounding IP~Per and BD$+$65$^\circ$1637.
When a central density enhancement is present, the radius at which the 
density profile merges with the field stars surface density provides 
an estimate of the cluster radius. The integral, I$_{\rm C}$, of the surface
density profile, subtracted of the field stars surface density determined 
at the edge of the images, gives an
estimate of the number of ``effective'' stars in the cluster.

\begin{figure}
\centerline{\psfig{figure=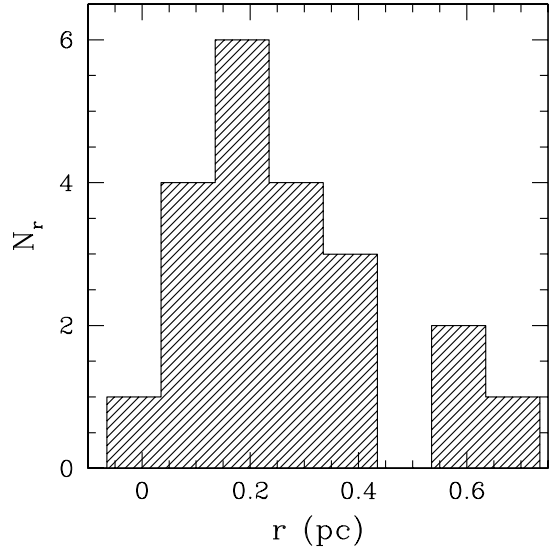,height=7.5cm,angle=0}}
\caption{\label{fr}Distribution of the cluster radii for the 21 stellar groups
detected with the source surface density profiles method}
\end{figure}

In Figure~\ref{fr} we show the distribution of the observed radii 
for the clusters
detected in the sample of Testi et al.~(1998). The distribution clearly shows
a peak for $r\sim$0.2~pc. This result is in good agreement with the typical
radii of young stellar clusters (Hillenbrand~1995; Carpenter et al.~1997). 
We note that this is the same size scale as the dense cores in molecular clouds,
which are the likely progenitors of stellar systems and clusters.

\begin{figure}
\centerline{\psfig{figure=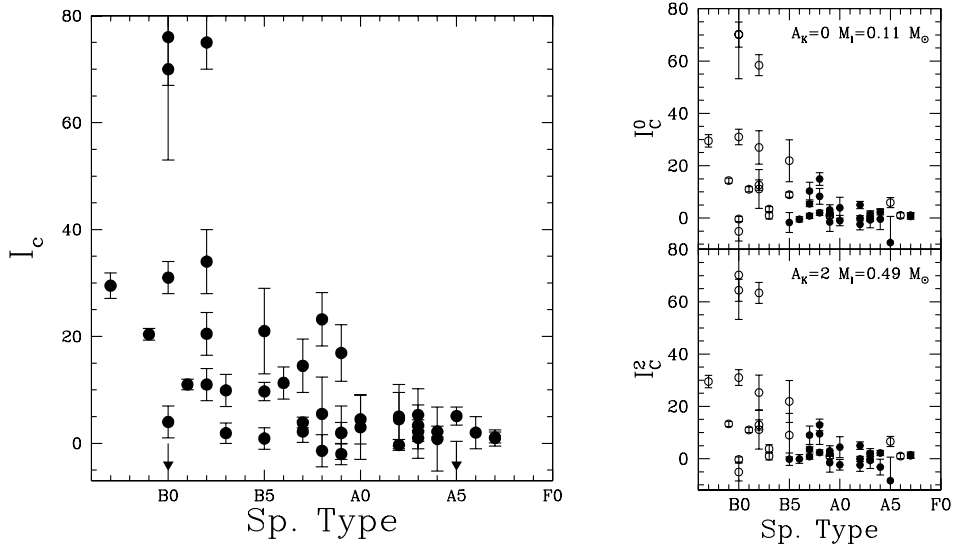,height=7.5cm,angle=0}}
\caption{\label{fic}Left: I$_{\rm C}$ vs. spectral type for the 44 Herbig Ae/Be
stars of our sample. Right: mass limited I$_{\rm C}$ vs. spectral type for 
A$_{\rm K}$=0 (top) and A$_{\rm K}$=2 (bottom).}
\end{figure}

The plot of I$_{\rm C}$ as a function of the spectral type of the target 
Herbig Ae/Be star (Figure~\ref{fic}) clearly shows the tendency for
earlier type stars to be surrounded by rich stellar groups. The same 
trend is evident when a correction for the different mass sensitivities in
the various fileds is applied (Figure~\ref{fic} right panels, see
Testi et al.~1999
for details). The lack of correlation between cluster richness and ages 
of the Herbig Ae/Be systems, together with the results of N-body simulations
suggest that different clustering properties of the stellar populations
around early-type Herbig Be stars (usually younger than $\sim$1~Myr) and
late-type Ae stars (usually older than $\sim$1~Myr) cannot be due to 
dynamical evolution but it is an imprint of the stellar formation mode. 

\section{Discussion}

In the previous section, we established that the observed correlation between
the spectral type of the Herbig Ae/Be stars and the membership number of the
stellar groups around them is not a result of observational or target sample
biases. The source of this correlation is thus to be searched in the star
formation process that produced the stars and the associated clusters.  In
particular, the results of our survey can be tested against the predictions
of two competitive classes of high-mass star formation models:  i) the
so-called {\it physical} models, that imply a physical relationship between
stellar clusters and massive stars, such as the coalescence models (Bonnell
et al.~1998; Stahler et al.~2000); and ii) the {\it random} sampling model,
in which stellar clusters are randomly assembled by picking stars from the
IMF: thus, massive stars are preferentially found in large stellar ensembles
because they are much less numerous than low-mass stars
(Elmegreen~1999;~2000; Bonnell \& Clarke~1999).

\subsection{{\it Physical} models}

The {\it physical} models imply that the mechanism of massive stars formation
is linked to the formation of a stellar clusters, or, in the extreme
view, the presence of a stellar cluster is a prerequisite for the 
formation of a massive star, such as in the coalescence models (Bonnell et
al.~1998; Stahler et al.~2000). These latter models require a very high
stellar density in order for (proto-)stellar merging to be efficient.

\begin{figure}
\centerline{\psfig{figure=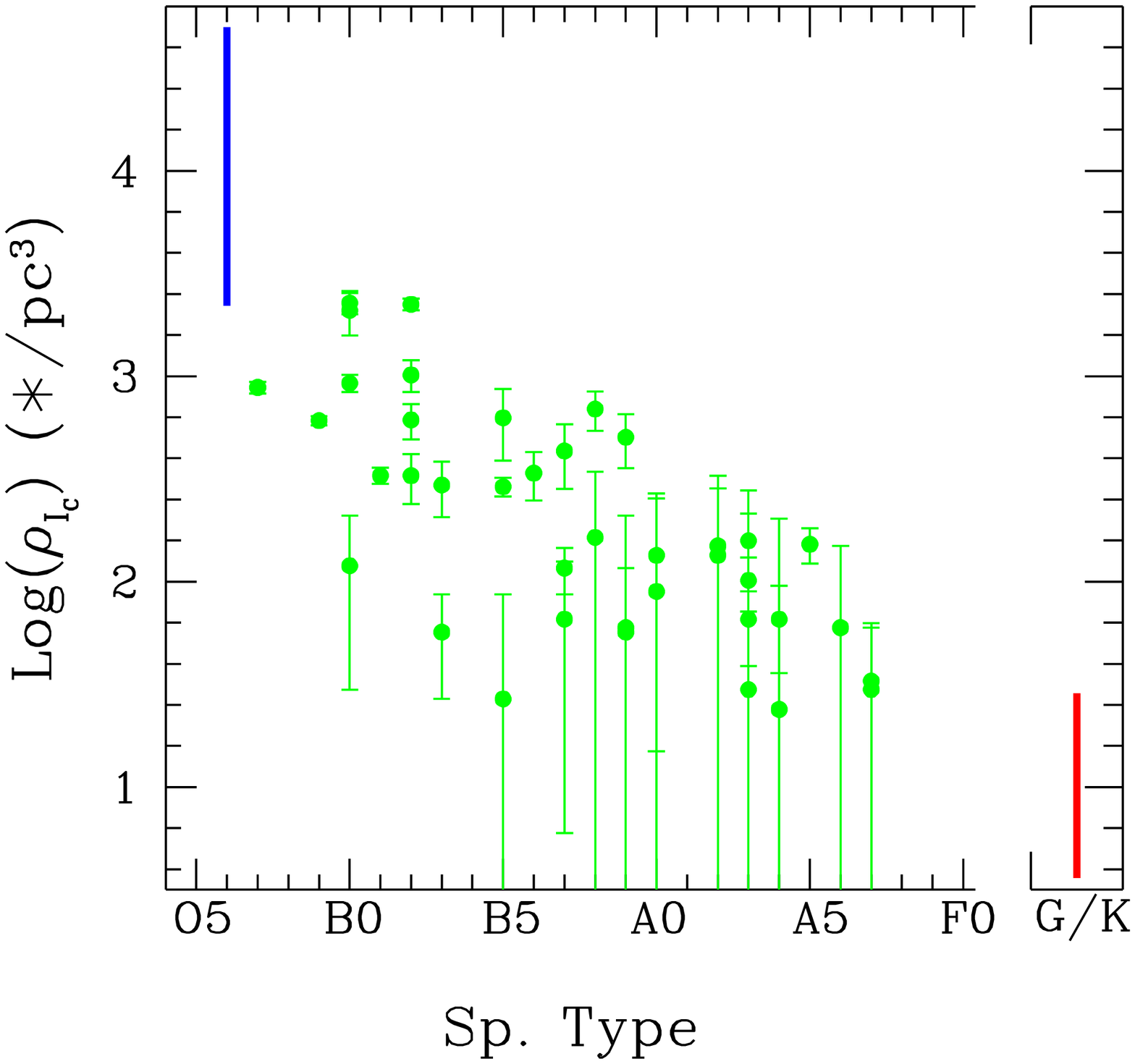,height=7.5cm,angle=0}
            \psfig{figure=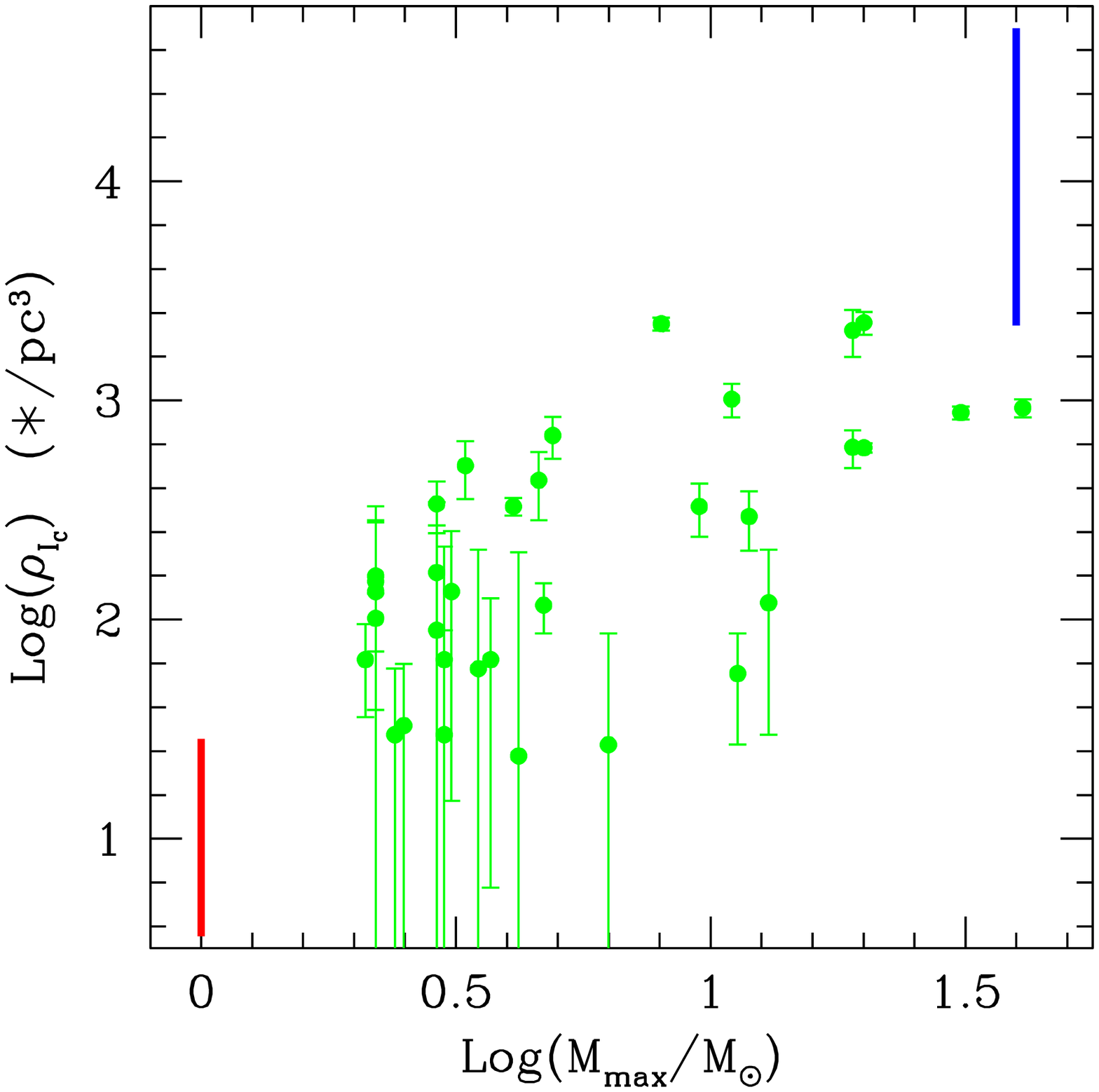,height=7.5cm,angle=0}}
\caption{\label{frho}Average stellar densities of the stellar groups detected
around Herbig Ae/Be stars versus spectral type (left panel) and
mass (right panel) of the target star. 
The solid lines show the typical densities of the young stars aggregates
in Taurus-Auriga (Gomez et al.~1993) and of the Orion Nebula Cluster
(Hillenbrand \& Hartmann~1998).}
\end{figure}

Given that all the detected clusters have approximately the same physical
size (see Figure~\ref{fr}), using the typical radius of $\sim$0.2~pc, the
number of effective stars can be converted into average stellar densities for 
spherical clusters. The result of this calculation (Figure~\ref{frho})
shows that while late-type Herbig stars are never surrounded by dense
stellar groups, the clusters around early Be stars reach densities in
excess of 10$^3$ stars per cubic parsec. 
The comparison with the typical densities of low mass stellar aggregates 
in Taurus-Auriga (Gomez et al.~1993) and that of the Orion Nebula Cluster 
(Hillenbrand \& Hartmann~1998) shows that the stellar groups detected around
Herbig Ae/Be stars nicely fill the gap between the two stellar density
regimes.
In Figure~\ref{frho} (right panel) we also show the average densities plotted 
against the mass of the most massive star in the cluster, i.e. the target
Herbig Ae/Be star. The masses of the stars have been computed from the 
data compiled by Testi et al.~(1998) using the 
pre-main-sequence evolutionary tracks of Palla \& Stahler~(1999) for
stars up to 6~M$_\odot$, and the zero-age main-sequence mass-luminosity 
relation of Tout et al.~(1996) for the most massive objects.
The plot clearly shows that only the most massive objects are associated 
with cluster densities $\gsim 10^3$~stars per cubic parsec.

This result appears to support the idea that interactions within a dense
cluster may play a role in the formation of the most massive stars, even
though one should be aware that the stellar densities that we derive are never
as high as those advocated by the Bonnell et al.~(1998) models.

\subsection{The {\it random} sampling model}

Bonnell \& Clarke~(1999) have shown that the observational data
are also consistent with the statistics of clusters with a defined 
membership size distribution randomly assembled picking stars from the
field stars initial mass function. In this view the formation of massive 
stars is not related to the formation (or the presence) of a dense
stellar cluster, but all stars are randomly selected as in
Elmegreen~(1999; 2000) models. The probability of finding
a cluster with N stars around a target star of a given mass, which by
definition will be the most massive star of the stellar ensemble, is
a function of both the stellar IMF and the clusters membership size
distribution g(N).
The latter is an unknown quantity since, at present, it is not
observationally constrained nor predicted by star formation theories.
\begin{figure}
\centerline{\psfig{figure=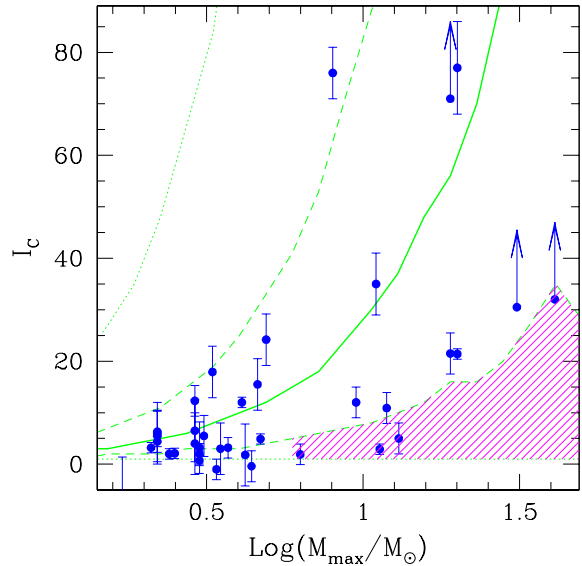,height=7.5cm,angle=0}}
\caption{\label{fic_ran}I$_{\rm C}$ vs. maximum stellar mass for the
Herbig~Ae/Be stars of our sample. The solid line shows the median values
expected for the {\it Random} model described in the text, the dashed lines
mark the first and third quartiles of the realizations.}
\end{figure}
Under the assumption that the probability of finding a cluster 
with membership size between N and N+dN ($\equiv \rm g(N)dN$)
is proportional to N$^{-\alpha}$, with 1.5$\le\alpha\le$2.0,
the {\it random} model predictions are compatible with the observed trend
and scatter of the I$_{\rm C}$ vs. maximum stellar mass as observed for 
the Herbig Ae/Be stars in the Testi et al.~(1999) sample (Bonnell \&
Clarke~1999). In Figure~\ref{fic_ran} we show the prediction of 
the {\it random} model against the Herbig Ae/Be sample results: 
the plotted model assumes the stellar IMF of Scalo~(1998), $\alpha$=1.7,
and the observations to be complete down to 0.2~M$_\odot$. In the Figure
the solid line shows the median of the random realizations,
the dashed lines show the first and third quartiles,
while dotted lines include all the realizations for each target maximum mass.
At the low M$_{max}$ end, the model always predicts small clusters,
while high-mass targets are predicted to be found as the most massive stars of
both small and large clusters. This is due to the steep power law distribution
of the cluster sizes: even though among low-N clusters those with a 
high-mass star as the most massive star are a minority, these are very
abundant with respect to the high-N clusters which have the same 
high-mass star as the most probable M$_{max}$ star. The probability
distribution of finding a cluster of given N around a selected mass
target show an increase of both the width and median values as a function
of the target mass.

It is interesting to note that the power law distribution
of the cluster membership sizes consistent with the observational data
is very close to the power law distribution of the mass spectrum of
gaseous clumps in molecular clouds (e.g. Blitz \& Williams~1999).
However, we believe that this correspondence is furtuitous and does not 
underscore any physical relationship between the quantites.
First, only a minor fraction of the observed clumps within a molecular
complex do form stars, the majority of them being gravitationally unbound.
Second, only some of the most massive clumps are bound, but there is no
indication that their mass distribution follows a power law with the
expected index. Obviously, it would be extremely interesting to verify
empirically the cluster membership distribution on a statistically
significant sample.

\subsection{How to discriminate between models}

The relationship between cluster stellar {\it density} and the
mass of the most massive star seems to favour the {\it physical} models,
while the increase in the scatter of the cluster richness for high-mass stars
and the presence of a few of these stars not surrounded by a conspicuous
stellar group appears to support the {\it random} model. However, none of
these two arguments is entirely convincing: average stellar densities have 
been derived assuming spherical clusters of radius $0.2$~pc, an
approximation that may not be valid in all cases; on the other hand,
the fact that few high-mass stars are not surrounded by a rich cluster can be 
explainded either in terms of observational problems (e.g. diffuse emission,
localised extinction, etc.) or by dynamical 
dissipation of the cluster wich is much more efficient for clusters
containing massive stars (see the discussion in Testi et al.~1999).

The best way to distinguish which of the two models is more 
appropriate requires a direct comparison with
a sample of high-mass target stars larger than that presently available.
While the {\it random} model predicts a relatively large fraction of massive 
stars in low-N clusters, the {\it physical} models posits that all 
massive stars are found within clusters, at birth time. More quantitatively,
the {\it random} model described in the previous section predicts that 
$\sim$25\% of the targets with mass exceeding $\sim$6~M$_\odot$ should be located
in the hatched area of Figure~\ref{fic_ran}, this prediction can be tested
by extending the survey of Testi et al.~(1998) to all 
of the early-type Herbig~Be stars catalogued in Th\'e et al.~(1994).
We are currently engaged in the completion of such a survey.
An additional test will be the search for stellar groups around the "field"
O-stars of Gies~(1987). These nearby O-stars are located away from known OB
associations and do not appear to have large proper motions that would
classify them as runaway stars. If these stars are really born in isolation
that would be against the {\it physical} models. As for the Herbig stars, 
they are relatively young, and, being optically visible, the search
for low-mass fainter companion is eased by the reduced extinction.

\begin{figure}
\centerline{\psfig{figure=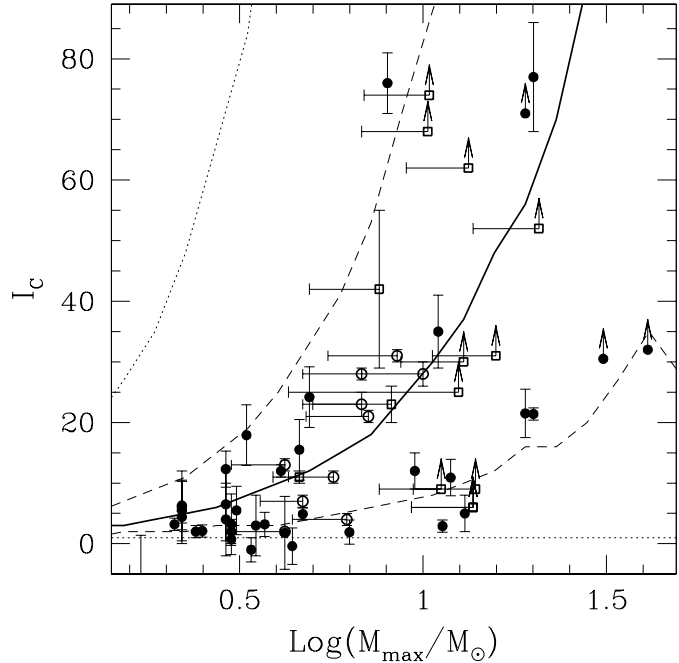,height=9.0cm,angle=0}}
\caption{\label{fic_comb}I$_{\rm C}$ vs. maximum stellar mass for the Herbig
Ae/Be stars sample (filled circles), the luminous IRAS Class~I sources in the
Vela-D GMC (open circles, Massi et al.~2000),and a sample of massive
protostellar candidates (open squares, Molinari et al.~2000).
The solid line shows the random model as in Fig.~\ref{fic_ran}.}
\end{figure}

An alternative approach will be to search for clusters around younger,
embedded high-mass sources. This method has the advantage of eliminating the
effects of dynamical evolution of the clusters, which is very difficult to
quantify properly (see Testi et al.~1999 and Bonnell \& Clarke~1999). A
number of NIR surveys of high-mass candidate Class~0 and Class~I sources are
being completed, and will be confronted with the predictions of the {\it
physical} and {\it random} models. The main problems in placing these objects
on the diagram of Fig.~\ref{fic_ran} are: i) the lack of a good estimate of
the maximum stellar mass, which is usually estimated from the total far
infrared luminosity that must be corrected for the contribution of low-mass
stars; ii) the fact that the star formation process is not yet completed in
these regions: thus, the clusters (if present) are expected to be deeply
embedded and not yet fully populated.  In any case, both effects tend to
overpopulate the region of the diagram at high M$_{max}$ and low cluster
richness. In Figure~\ref{fic_comb} we show the preliminary results of two of
these surveys (Massi et al.~2000; Molinari et al.~2000) together with the
Herbig~Ae/Be sample. Although these surveys do not include objects with
M$_{max}\gsim 20$~M$_\odot$ (or L$_{bol}\gsim$5$\times$10$^4$~L$_\odot$),
these initial results indicate that the {\it random} model predicts a higher
proportion of low membership clusters.  However, a more firm conclusion has
to wait for the completion of these surveys, especially at the high-mass
end.

\section{Conclusions}

The results of a survey for young embedded clusters around
a sample of 44 Herbig Ae/Be stars show a clear dependence
of the cluster richness on the mass of the Herbig Ae/Be stars,
the most massive object of the group.
Stars of progressively higher mass appear to be surrounded by richer and
denser clusters. These findings are in qualitative agreement with
models that suggest a causal relationship between the birth of a massive
star and the presence of rich stellar clusters (the
{\it physical} model). The observed
correlation and scatter, however, could also be explained in terms
of random assembling from a standard IMF picked from a 
membership size distribution of the form g(N)$\sim$N$^{-\alpha}$, with
1.5$\lsim\alpha\lsim$2.0 (the {\it random} model). 
We have outline several ways of discriminating 
between these two competitive models. For the moment, the jury is still out.

\end{document}